\documentclass[a4paper,11pt]{article}
\pdfoutput=1 % if your are submitting a pdflatex (i.e. if you have
\usepackage{appendix}
\usepackage{enumitem}
\usepackage{color}
\usepackage[normalem]{ulem}
\usepackage{amssymb}
\usepackage{jinstpub} % for details on the use of the package, please
\title{Development and implementation of a time-based signal generation scheme for the Muon Chamber simulation of the CBM experiment at FAIR}

\author[a,b,1]{V.~Singhal,\note{Corresponding author.}}
\author[c]{S.~Chatterjee,}
\author[d]{V.~Friese,}
\author[b]{S.~Chattopadhyay}

\affiliation[a]{Homi Bhabha National Institute, Kolkata, India}
\affiliation[b]{Variable Energy Cyclotron Centre, 1/AF Bidhan Nagar, Kolkata-700 064, India}
\affiliation[c]{Department of Physics (CAPSS), Bose Institute, Kolkata - 700091}
\affiliation[d]{GSI Darmstadt, D-64291 Darmstadt, Germany}

\emailAdd{vikas@vecc.gov.in}

\abstract{
The Compressed Baryonic Matter~(CBM) experiment in the upcoming Facility for Antiproton and Ion Research~(FAIR), designed to take data in nuclear collisions at very high interaction rates of up to 10 MHz, will employ a free-streaming data acquisition with self-triggered readout electronics, without any hardware trigger. A simulation framework with a realistic digitization of the detectors in the muon chamber (MuCh) subsystem in CBM has been developed to provide a realistic simulation of the time-stamped data stream. In this article, we describe the implementation of the free-streaming detector simulation and the basic data related effects on the detector with respect to the interaction rate.
}

\begin{document}
\maketitle
\flushbottom

\section{Introduction}
\label{sec:intro}

The data acquisition systems of most experiments in high-energy particle or nuclear physics are based on a hardware trigger, where a signal generated by a suitable set of hardware indicates that a collision took place and triggers the timely readout of the front-end electronics. The hardware trigger thus defines ``event'' as data collections of moderate size representing a separate single collision, which may then be either written to permanent storage or subjected to further inspection and selection by higher-level triggers, e.g., on FPGA or in software. The software framework used for simulation and analysis of such experiments are thus designed on an event-by-event scheme, where each event is treated as a separate and independent entity.

Several next-generation experiments, however, face the difficulty that such a triggered readout scheme is not feasible, e.g., because of very high interaction rates and/or complex trigger topologies which are not well suited to be evaluated in hardware logic and time consuming. An example is the Compressed Baryonic Matter experiment (CBM)~\cite{cbm} at the upcoming Facility for Antiproton and Ion Research (FAIR)~\cite{fair} in Darmstadt, Germany. CBM intends to inspect up to $10^7$ nuclear collisions per second, each producing several hundreds of particles to be registered in the detectors. These conditions will use a trigger-less, free running data acquisition, where self-triggered front-end electronics elements register signals above threshold caused by particles traversing the respective detectors and autonomously push the data forward~\cite{TriggerLess}. Such a system is not limited by latency, i.e. the time needed to generate a hardware trigger, but by data throughput bandwidth. It results in a continuous data stream in contrast to a series of events defined by the hardware trigger in conventional readout schemes. The association of raw data to physical events is based on a precise time stamp provided to each measurement by a central timing system.

The software to analyse this data stream - both in real-time and offline - must cope with this situation~\cite{friese2012}. The same holds also for the software to simulate the experiment, which must generate such a data stream from appropriate physics models of high-energy nuclear collisions. The simulation thus has to convert a series of events into a stream of data incorporating response of detectors, front-end electronics and data acquisition to these events. The CBM software framework cbmroot~\cite{cbmroot}, based on the FairRoot simulation and analysis framework~\cite{FlorianFairRoot}, provides the appropriate structures for this task, in particular the software emulation of the data acquisition collecting the free-streaming data from the detector front-ends~\cite{friese2017}. These framework structures must of course be filled by the implementation of the actual detector response of the various CBM sub-systems. 

In this article, we describe the time based simulation of the CBM muon detector system MuCh. In section~\ref{sec:cbm_much} the CBM experiment and its muon detector are introduced. Section~\ref{sec:digitizer} then describes the implementation of the detector response for the MuCh detector. Verification and performance of the simulation are discussed in section~\ref{sec:results}. Finally, section~\ref{sec:summary} gives a summary and conclusions.

\section{The CBM experiment and its muon detector}
\label{sec:cbm_much}

\begin{figure}[htb]
\begin{center}
\includegraphics*[width=0.7\linewidth]{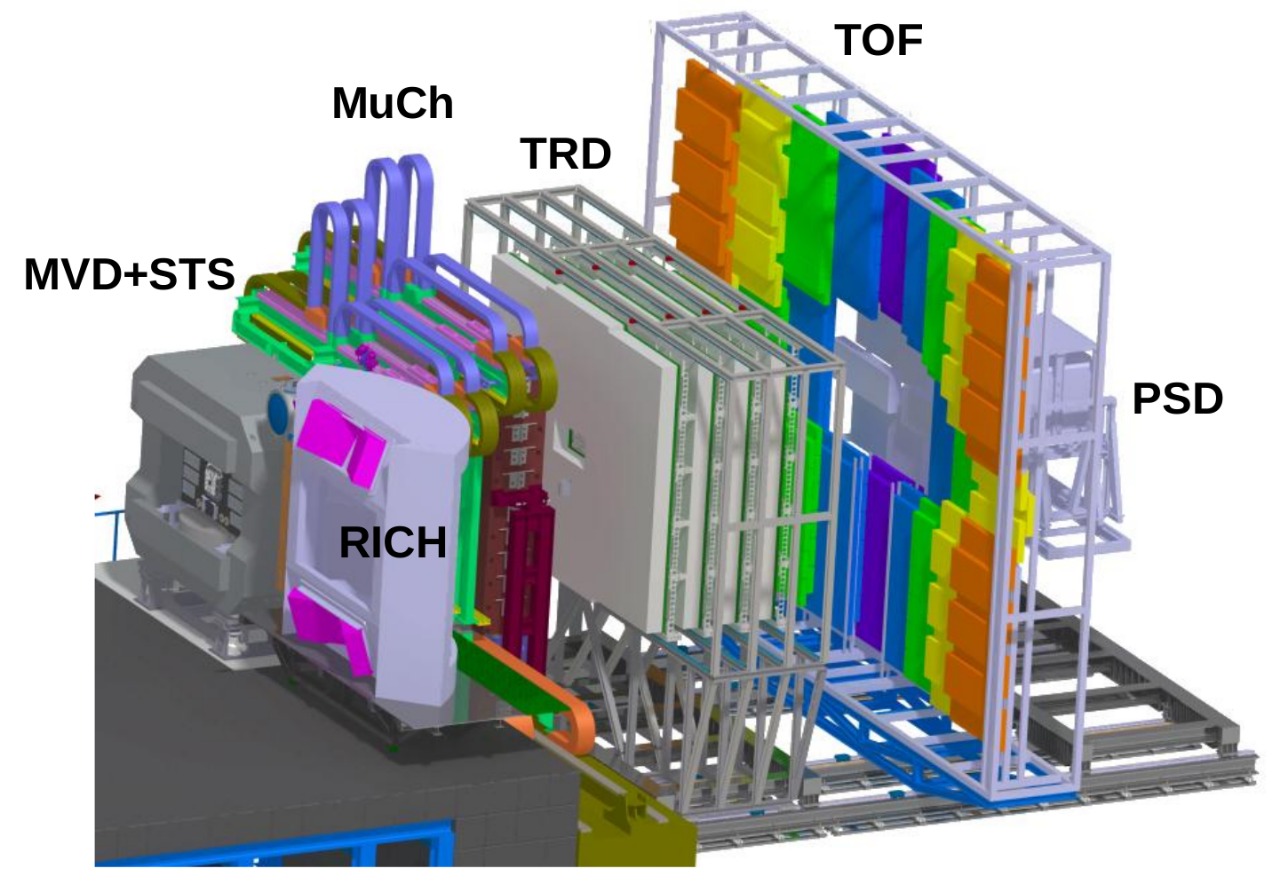}
\caption{Layout of the CBM experiment in its muon configuration. RICH will be in parking position during MuCh dating data. The beam enters from the left. The magnet houses the target, MVD and the central tracking system STS. The muon system is placed directly after the magnet, outside of the magnetic field. Further downstream are the TRD, TOF and PSD detectors. }
\label{fig:muon_setup}
\end{center}
\end{figure}

CBM is a fixed-target experiment for the study of energetic nuclear collisions at beam momenta from 3.5 -- 12 GeV per nucleon for heavy nuclei~\cite{cbmpaper}. It will be operated with external beams from the FAIR accelerator complex from 2025 on. CBM comprises two detector configurations: a setup for measurements of hadrons and electrons, with a RICH and a TRD detector for electron identification and a TOF detector for hadron identification, and a setup for measurements of muons, in which the RICH detector is replaced by the muon system MuCh. Both configurations share the central tracking system STS inside the field of a superconducting dipole magnet, and the forward calorimeter PSD. The muon configuration of CBM, with RICH in parking position, is shown in Fig~\ref{fig:muon_setup}.

\begin{figure}[htb]
\begin{center}
\includegraphics*[height=0.45\linewidth]{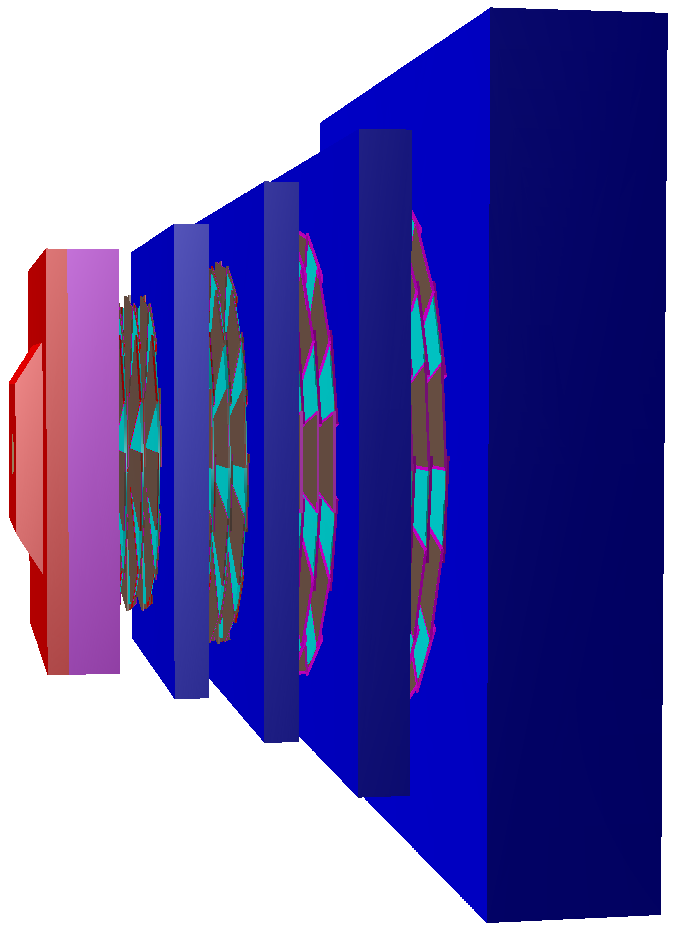}
\hfill
\includegraphics*[height=0.45\linewidth]{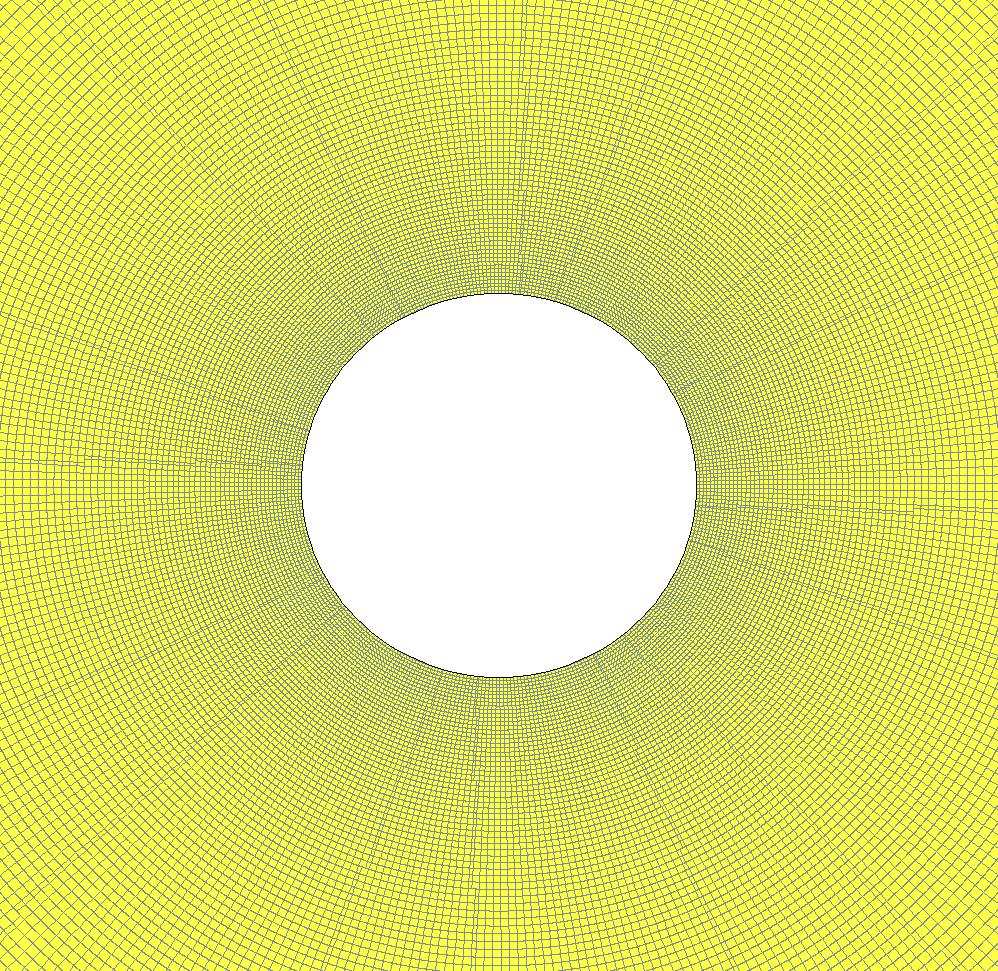}
\caption{Left: Schematic layout of the MuCh detector, consisting of a set of absorber segments inter-laid with detector stations. Right: Pad segmentation of the first layer of the first station. $1^{\circ}$ in  $\phi$, \& $\delta$r $=$ $r\delta\phi$.}
\label{fig:much_geometry}
\end{center}
\end{figure}

The MuCh detector~\cite{muchwiki, muchtdr} serves for the identification of muons by filtering out other particle species in massive absorbers. The absorbers are arranged as five segments with three detector layers in between each pair of segments. The segmented absorber system allows to trace particle trajectories through the entire setup, which would not be possible by a monolithic absorber because of small-angle scattering of the muons in the absorber material. A thick absorber also likely to absorb muons without generating signal. Figure~\ref{fig:much_geometry} (left) shows the MuCh geometry as implemented in cbmroot.

The hit rates in the detector layers differ significantly because of the successive absorption of particles by the absorber segments. Thus, different detector technologies are employed: the two upstream stations (detector triplets) will use Gas Electron Multiplier (GEM detector), while the two downstream stations will be constructed from high-rate Resistive Plate Chambers (RPC). In both cases, the detector readout planes are segmented into pads in a $r-\phi$ geometry, the granularity depending on the layer position and the radial distance from the beam. The readout segmentation was optimised based on efficiency and signal-to-background ratio for the detection of muon pairs in heavy-ion collisions at FAIR energies~\cite{muchtdr, segmentation}. The pad layout is illustrated in Fig.~\ref{fig:much_geometry} (right) on the example of the first layer of the first station.

\section{Detector response simulation}
\label{sec:digitizer}

\subsection{Simulations in cbmroot}

In cbmroot, simulations are performed in two separate steps with intermediate file output. In the first step called ``transport simulation‘‘, the primary particles are traced through the detector, taking into account their trajectory in the magnetic field, their interaction with the materials and - for unstable particles - their decay. Secondary particles created by decay or interaction with materials are created and traced as well. This step employs external engines; the framework makes use of the ROOT TVirtualMC features which allow to choose between GEANT3 and GEANT4 on the discretion of the user~\cite{vmc}, however, GEANT3 is used for all the results mentioned in this article. At transport level, a realistic description of the detector geometries, the proper material properties and a map of the magnetic field are required~\cite{MuchGeometry}. As output of the transport simulation, the geometric intersections of particle trajectories with active detector elements (``MCPoints'') are recorded, along with the time-of-flight from the event start and the energy deposit in the active material. This information is stored to file and provides the input for the second simulation step.

The task of the detector response simulation is to model the physics processes in the active detector material, the readout and the digitization in the front-end electronics. The output objects are called ``digi'' represent the simulated measurements of a single read-out channel - corresponding to one read-out pad in the case of the MuCh. This atomic data unit contains the information on the respective channel address, the time of the measurement termed as time-stamp, and the digitized charge. The digi objects are then forwarded to the software emulation of the CBM DAQ system as implemented in cbmroot. At this step, the association of digis to events is destroyed, resulting in a data stream similar to that expected from the actual experiment. In the time-based simulation, based on the dead-time discussed later, a single pad might be recorded as two digis with different time-stamp. 

In the context of a free-streaming readout as described in section~\ref{sec:intro}, the correct description of the timing behaviour is of utmost importance. The effects contributing to the finally registered time stamp are:
\begin{itemize}
    \item Event time: the time corresponding to the actual collision. It is generated by the cbmroot framework from the time profile of the beam and the interaction probability of the beam particles in the target.
    \item Time-of-flight: from the event start to the intersection with the detector element. This time is provided by the transport engine and stored in the MCPoint object.
    \item Drift time: the time taken by the primary electrons in the GEM or RPC detectors to drift to the readout planes. 
    \item Time response of the readout ASIC to the analog charge collected in the readout pads.
\end{itemize}

An important figure is the dead time of the readout electronics, i.e., the time after a hit in which the corresponding readout channel is blocked, such that a second hit arriving within this time is neglected. This leads to a detector inefficiency and track pile-up which are obviously dependent on the interaction rate, i.e., the time separation between two subsequent collisions.

\subsection{Analogue response simulation}

\begin{figure}[htb]
\begin{center}
\includegraphics*[height=0.3\linewidth]{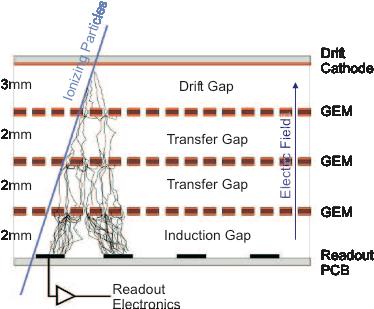}
\hfill
\includegraphics*[height=0.3\linewidth]{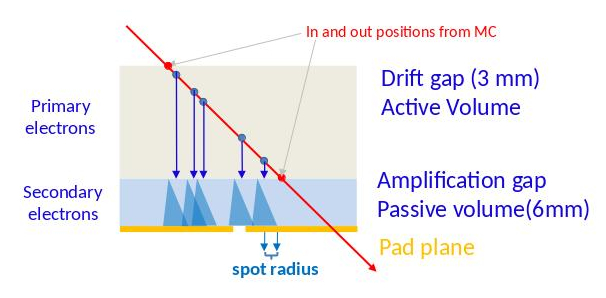}
\caption{Left: working principle of a multi-layer GEM detector; Right: simplified scheme as implemented in the analogue response simulation}
\label{fig:analog_response}
\end{center}
\end{figure}

The working principle of a multi-layer GEM detector is illustrated in Fig.~\ref{fig:analog_response}. A traversing charged particle creates a number of primary electrons in the drift gap through ionisation of the gas. The primary electrons drift in the electrical field towards the GEM foils, where amplification through the creation of avalanches takes place. This process is repeated in various transfer gaps between successive GEM foils. Finally, the produced electrons are further amplified in the induction gap and collected on the pads of the readout PCB.

Our analogue simulation implements a simplified scheme as shown in the right panel of Fig.~\ref{fig:analog_response}. The trajectory of the charged particle in the drift gap is approximated by a straight line, the coordinates of the entry and exit points being provided from the transport simulation stage. For each particle, the number of primary electrons is sampled from a Landau distribution, the parameters of which depend on particle type, energy, track length in the active volume and specifications of the gas mixture~\cite{gemparameters}. The primary electrons are randomly generated along the trajectory in the active volume with a uniform distribution.

For each primary electron arriving at the amplification gap, the number of secondary electrons is calculated from the gas gain settings, and the charge depositions to the gaps are calculated assuming a spot radius of the avalanche. Depending on the track inclination, the gas gain and the avalanche profile, a particle can activate one or several pads. The drift time is the time taken by the primary electron to traverse the active volume (up to the first GEM foil). The avalanche travels with close to the speed of light, such that its propagation time is negligible in comparison to the drift time. Primary electrons generated in the passive volume do not contribute measurably to the total signal since no avalanche production takes place.

The parameter input to this analogue simulation consists of the parameters of the Landau distribution for the number of primary electrons, the drift velocity of the primary electrons, the gas gain, and the avalanche spot radius. All of these parameters are tunable in the software. The size of the drift gap is taken as per geometry of GEM or RPC. The analogue simulation process is the same for GEM and RPC detectors, but with different parameters. The default settings for both detector types are summarised in Table~\ref{tab:analog_parameters}.
 
\begin{table}[h]
    \begin{center}
        \caption{\label{tab:analog_parameters}Parameters for the analogue simulation of GEM and RPC detectors}
        \begin{tabular}{l|l|l}
        \hline
        Parameter & GEM & RPC \\
        \hline
        MPV for primary Landau distribution & from HEED & 12 \\
        Drift gap (fixed as per detector geometry)  & 3 mm & 2 mm\\
         Drift Velocity of primary electrons ($\mu$m/ns)& 100 & 120\\
        Gas gain & 5000 & 30000\\
        Spot Radius & 500 $\mu$m & 2 mm\\
        \hline
        \end{tabular}
    \end{center}
\end{table}

The analogue simulation as described above is performed separately for each input MCPoint. 
The resulting charge depositions to the readout pads, however, cannot directly be digitised, since they block the respective electronics channel for a certain amount of time (dead time), thus potentially influencing later measurements. Such incidences of different particles contributing charge to the same pad can happen within one event (conventional pile-up), but in our free-running scenario also for particles from different events, provided the difference in the event times is comparable or smaller than the dead time. To cope with this situation, the registered charges per pad are internally buffered and keeps a match object associating a link per corresponding incident track. Later this match information is used for finding the primary and secondary track contribution towards generation of digi or pile-up. The stored information also contains the pad address, the signal time, the charge, and the duration of the active signal.

\subsection{Digital response simulation}

The digital response to the accumulated charge in the readout pads is modelled following the properties of the readout ASIC, the STS/MUCH-XYTER (SMX)~\cite{iaea,nxyter}. This ASIC features two shaping channels, the fast shaper used to determine the time stamp of the measurement, and the slow shaper which measures the charge amplitude by time-over-threshold. Once a signal in the fast shaper crosses a pre-defined threshold, the signal in the slow shaper is evaluated and a message to the DAQ system is issued when the amplitude in the slow shaper falls below a second threshold. The signal shape in the slow shaper thus defines under which conditions two subsequent signals in the same readout channel can be disentangled.

This situation is modelled in our software by a delta function for the signal shape, the start time being the time stamp and the duration being the dead time of the ASIC. The amplitude is the total charge. Two overlapping signals (i.e., the second one arriving within the dead time of the first) are merged into one, having the start time of the first signal, the stop time of the second, and the amplitude the sum of both signals. The generated signals are transiently stored in a readout buffer. When the system time (the time of the currently processed event) exceeds the stop time of the signal, it is digitised, since then it is guaranteed that it will not be influenced by subsequently arriving signals. 

All analogue signals exceeding the threshold are digitised, using the parameters of the ADC built into the SMX. The charge is linearly discretized, above a tunable threshold, into 32 bins corresponding to the 5 bit resolution. The time of the digital signal is smeared by a Gaussian distribution representing the time resolution of the ASIC. The parameter input to the digital response simulation are thus the number of ADC channels, the ADC threshold and dynamic range, and the time resolution. The currently used values of these parameters are summarised in Table~\ref{tab:digital_parameters}. It should be noted that these parameters are still subject to changes; in particular, the dead time, which is a critical parameter for the performance of the detector system, is expected to be smaller for SMX v2.2 currently under development~\cite{SMX}.

\begin{table}[h]
    \begin{center}
        \caption{\label{tab:digital_parameters}Parameters for the digital response simulation of GEM and RPC detectors}
        \begin{tabular}{l|l|l}
        \hline
        Parameter & GEM & RPC \\
        \hline
        Number of ADC channels& 32 (5 bit) & 32 (5 bit)\\
        ADC threshold    & 2 fC & 30 fC \\
        ADC dynamic range  & 80 fC & 130 fC \\
        Dead time   &  400 ns   &   400 ns  \\
        Time resolution  &  5 ns  &  5 ns \\
        \hline
        \end{tabular}
    \end{center}
\end{table}

\subsection{Noise simulation}

Our simulation also covers thermal noise generating random signals from the readout electronics. Thermal noise charge is always present in the readout channel; the distribution of its value is approximately Gaussian. Whenever the noise charge exceeds the threshold, a message is generated just as in case of charge originating from a traversing particle. The occurrence of this noise is random and not correlated in time to events.

In our implementation, the noise is described by a tunable parameter, the noise rate per pad. From this parameter, the number of noise signals is sampled for each module for the time period from the previous to the current event time from a Poissonian distribution with the expectation value noise rate $\times$ number of pads per module x time interval. The time of the noise signals is sampled from a uniform distribution in the respective time interval; their charge is sampled from the Gaussian noise distribution. 

Noise signals are inserted into the readout buffer and thus into the simulated data stream. They can interfere with ''real`` signals from traversing particles in the same way as two ''real``signals (see the previous subsection).

\section{Results and detector performance}
\label{sec:results}

\begin{figure}[htb]
\begin{center}
\includegraphics[width=0.8\linewidth]{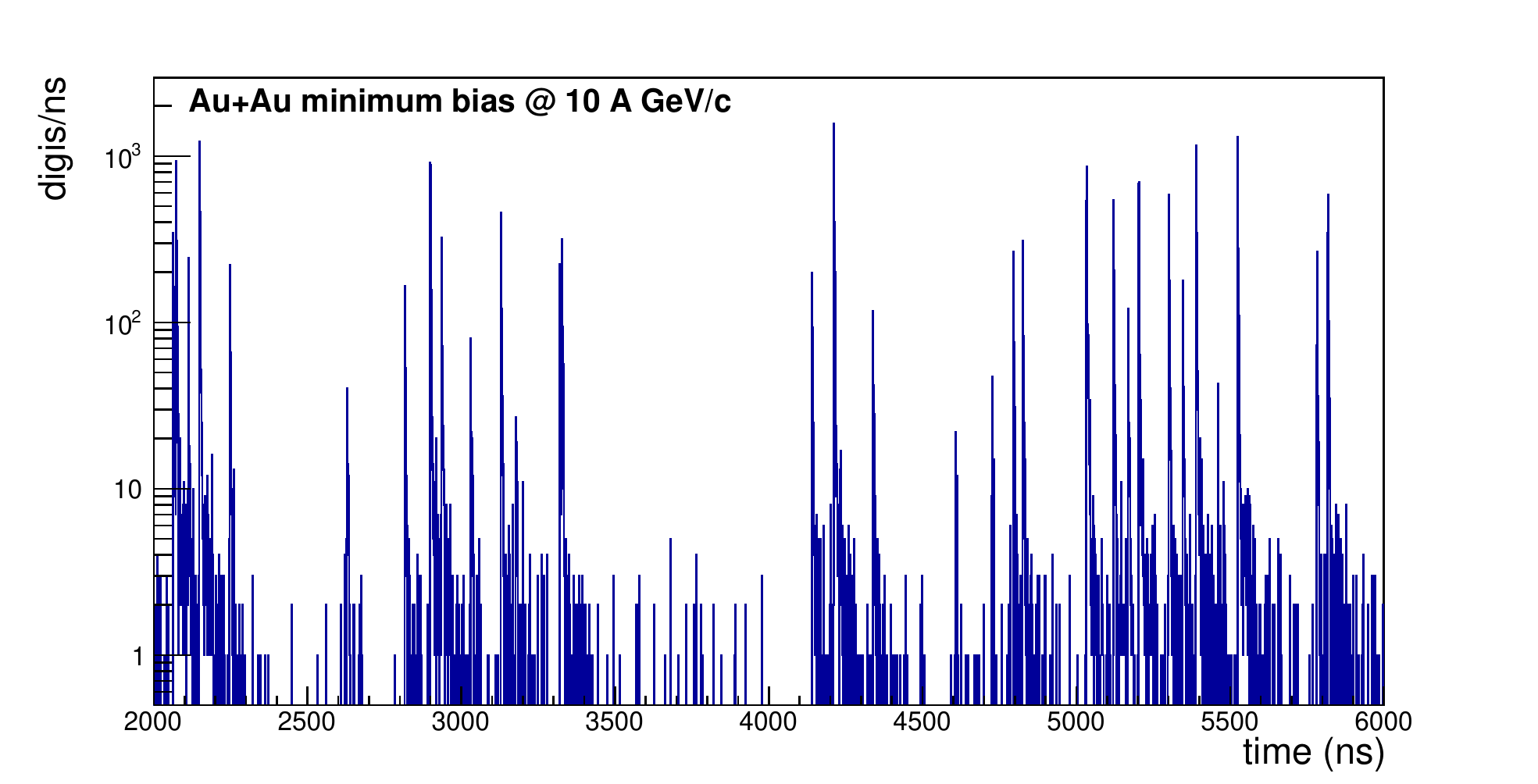}
\caption{The time distribution of the MuCh digis for a certain time period ( within a time slice), with several events in and with moderate noise switched on, resembling self-triggered free running data stream.}
\label{fig:datastream}
\end{center}
\end{figure}

The result of the MuCh simulation is a stream of digi objects integrated by the DAQ software into the streams of the other CBM detector systems. Figure~\ref{fig:datastream} shows the digi rate obtained from the simulation. The MC-true event time generated by the cbmroot simulation framework is overlaid for comparison. We see that events can be identified by peaks in the distribution of digis~\cite{eventbuilding}. 

\begin{figure}[htb]
\begin{center}
\includegraphics[width=0.8\linewidth]{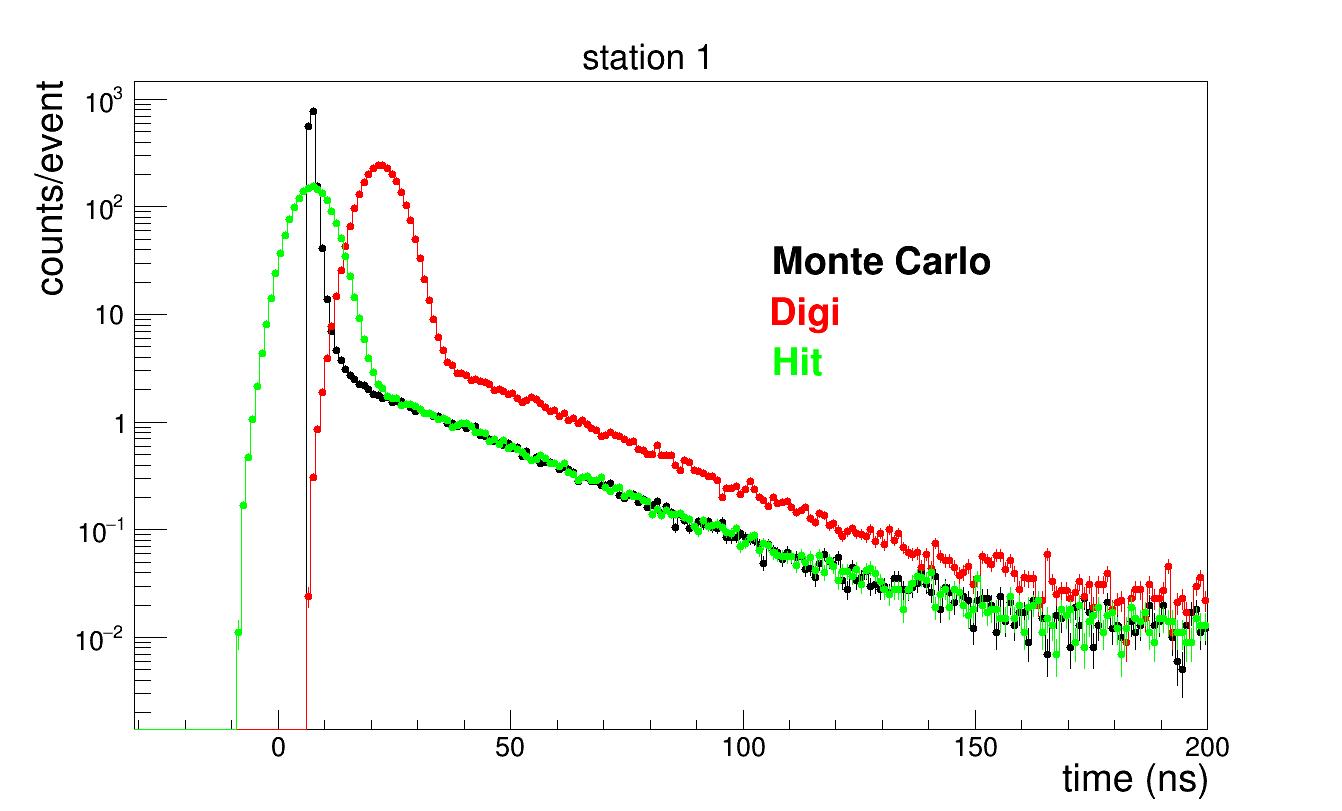}
\caption{Time distribution of Monte-Carlo points, digis and hits in a single event in the first station of the MuCh detector. Hits are reconstructed from clusters of digis in neighbouring active pads. The origin of the time axis corresponds to the Monte-Carlo event time.}
\label{fig:time_event}
\end{center}
\end{figure}

The time distribution of digis within a single event is shown in figure~\ref{fig:time_event} together with the corresponding Monte-Carlo true time of the track in the detector. Included is also the reconstructed hit time, where a hit results from a cluster-finding procedure which groups simultaneously active neighbouring pads. The MC true time shows a narrow peak and an approximately exponential tail. The peak originates from both primary and secondary tracks, which are registered within a time span of less than 10 ns. Due to interaction with absorber material of muon system, 94.5\% digis, registered in MuCh detector layers, are generated from secondary tracks. The offset of the peak (about 7 ns) reflects the time-of-flight from the main interaction vertex to the MuCh detectors. The exponential tail comprises secondary tracks only. In comparison, the digi time distribution shows a broadening of the peak due to the time resolution and an additional offset stemming from the drift time of the electrons in the drift gap of the GEM counters (see Fig.~\ref{fig:analog_response}). Note that the integrals of the distributions of MC points and digis are different since one track can activate more than one pad and more than one MC points can generates one digi. The hit time, on the other hand, is corrected for the electron drift and thus shows no offset with respect to the MC origin. 

The CBM experiment intends to operate at high interaction rates of up to $10^7$ events per second. At such rates, the average time between two subsequent events (100 ns for $10^7$ events/s) becomes comparable to the dead time of the readout electronics (400 ns). Consequently, losses due to pile-up from particles originating from different events are expected. The developed simulation software as described in the previous section allows to study such timing effects and to quantify rate-dependent losses, which is essential to determine the performance of the detector. 

For the following studies, we simulated both minimum-bias and central Au+Au collisions at 10 GeV/$c$. In the minimum-bias event sample, the collision impact parameter $b$ is realistically distributed according to the geometric cross section ($\propto b\mathrm{d}b$). The time sequence of such events corresponds to the actual experimental situation. Central events ($b=0$) create the highest track multiplicity and thus the highest data load on the detectors. A time series of central events only does not correspond to the physical situation, but is used here to verify the simulation software since it magnifies effects arising from the finite dead time.

\begin{figure}[htb]
\begin{center}
\includegraphics*[width=0.6\linewidth]{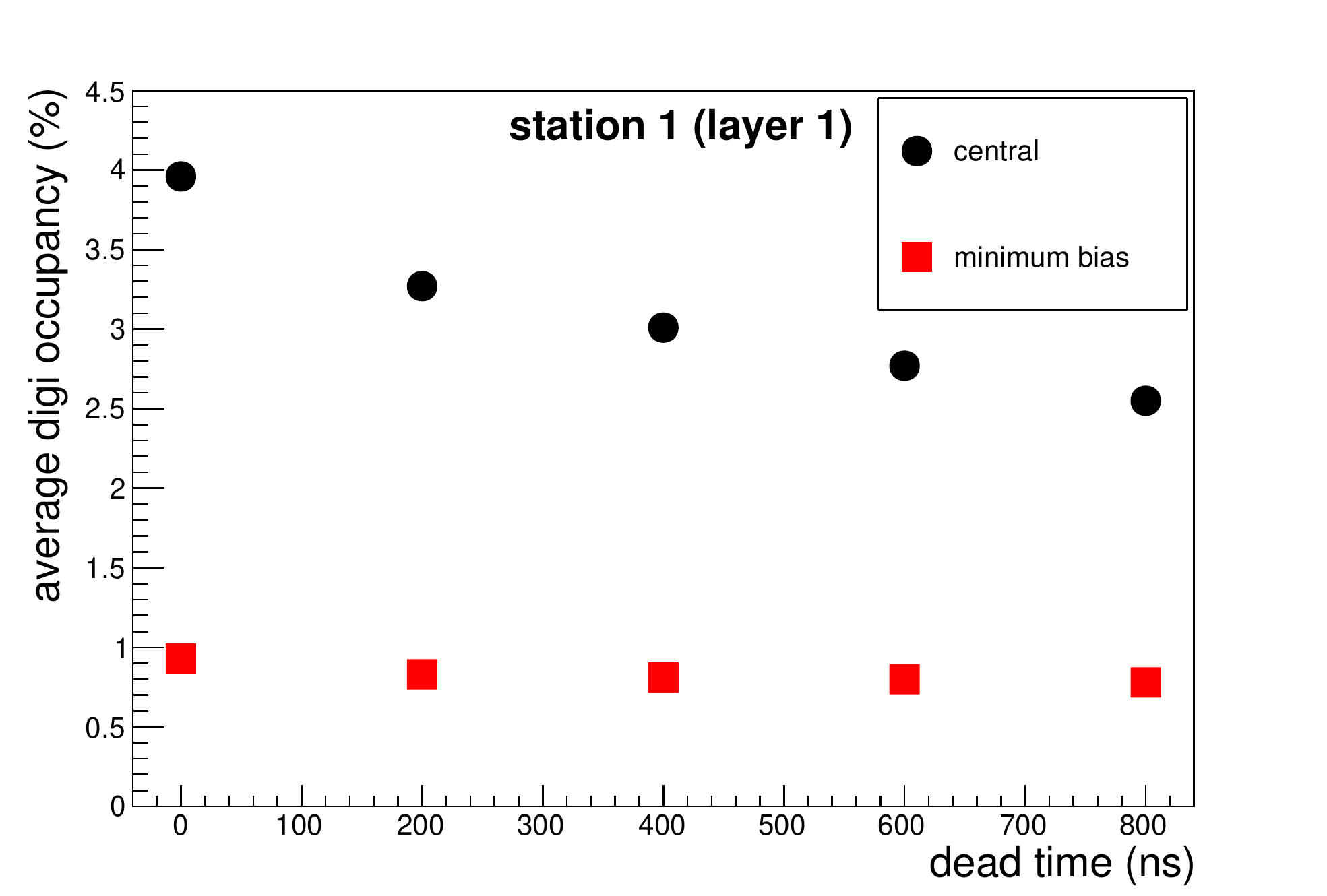}
\caption{Average single-channel occupancy in the first layer of the first station as function of dead time for Au+Au collisions at $10^7$ events/s}
\label{fig:occupancy}
\end{center}
\end{figure}

A common detector parameter figure is the occupancy, conventionally defined as the probability for a single channel to be activated in one event. In our free-running scenario, we calculate this quantity as the number of digis divided by the number of pads and the number of simulated events. Figure~\ref{fig:occupancy} shows this occupancy (averaged over one detector layer) for the first layer of the first station at an event rate of $10^7$/s as function of the single-channel dead time. The decrease of occupancy with dead time, moderate for minimum-bias events and better visible for central events, is an expected consequence of the data loss due to signals arriving in the dead time of a previous signal. 

\begin{figure}[htb]
\begin{center}
\includegraphics[width=0.6\linewidth]{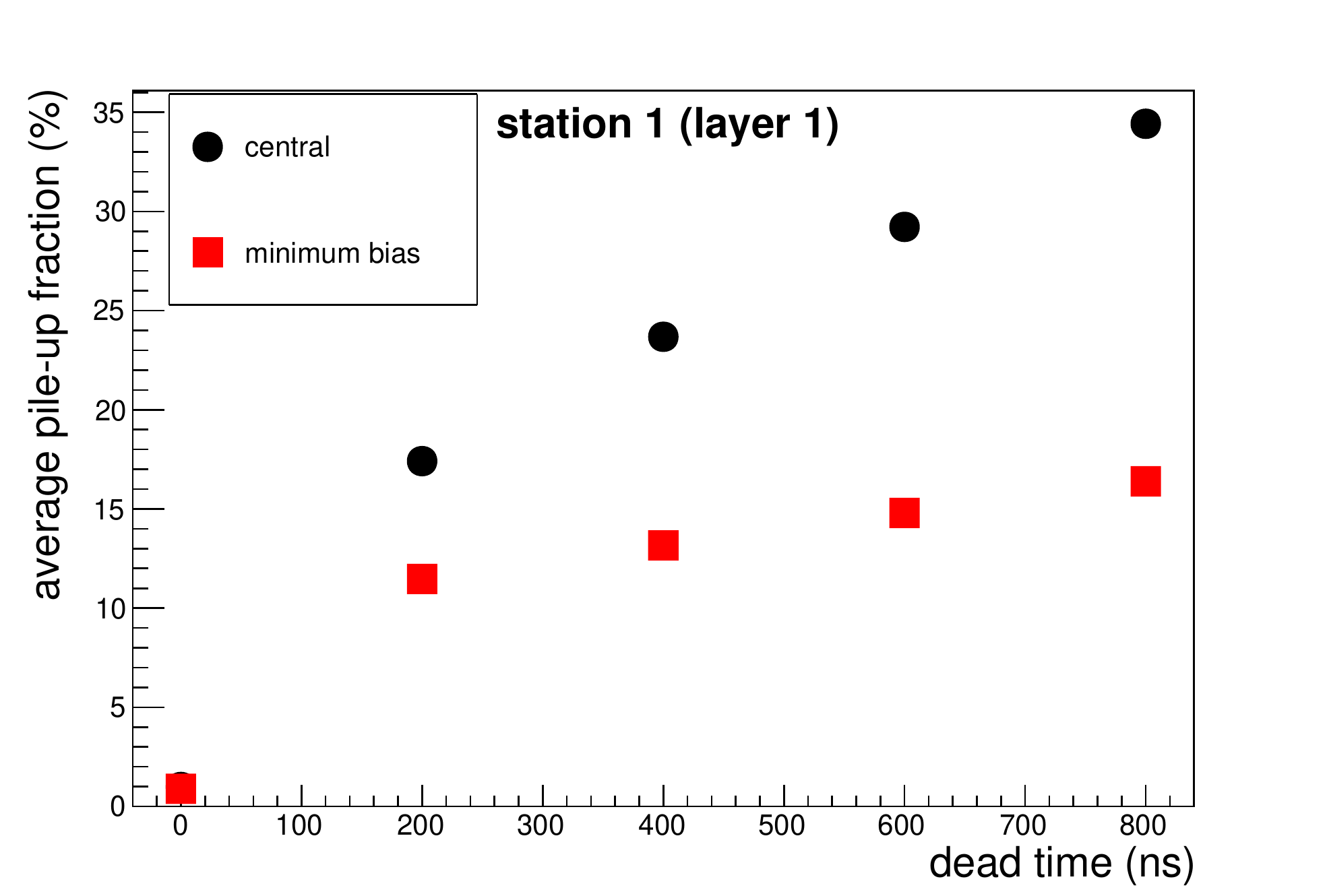}
\caption{Average pile-up fraction in the first layer of the first station as function of dead time for Au+Au collisions at $10^7$ events/s.}
\label{fig:pileup_deadtime}
\end{center}
\end{figure}

Rate-dependent data losses can also be qualified by studying the pile-up occurrences. In our simulation as earlier mentioned, the reference to the Monte-Carlo origin is kept for each digi object. Pile-up thus happened if a digi object has references to more than one Monte-Carlo particles. We define the pile-up fraction as the number of digis with multiple MC reference divided by the total number of digis. This fraction is shown in Fig.~\ref{fig:pileup_deadtime} again as a function of dead time for an interaction rate of $10^7$/s.  For minimum-bias events, the fraction slowly varies with the dead time; for a value of 400 ns, corresponding to the current SMX design, it amounts to about 13\%. It should be mentioned that as discussed later the maximum contribution to this pile-up fraction is from the secondary tracks generated in the material. This does not represent the pile-up for all tracks (primary \& secondary) associated with the incident primaries, which will govern the capabilities of tracking of primary tracks in the detector.

\begin{figure}[htb]
\begin{center}
\includegraphics[width=0.49\linewidth]{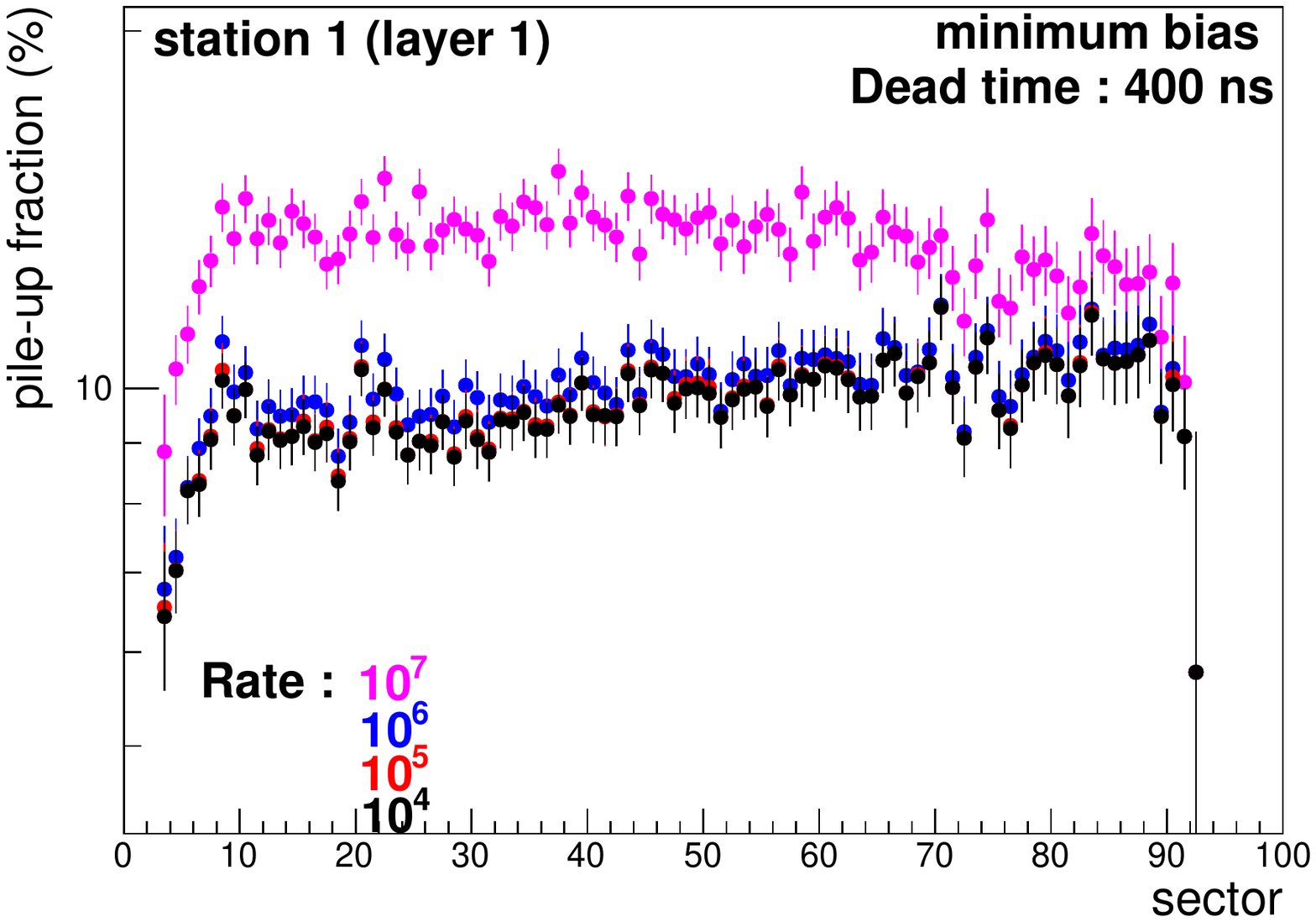}
\hfill
\includegraphics[width=0.49\linewidth]{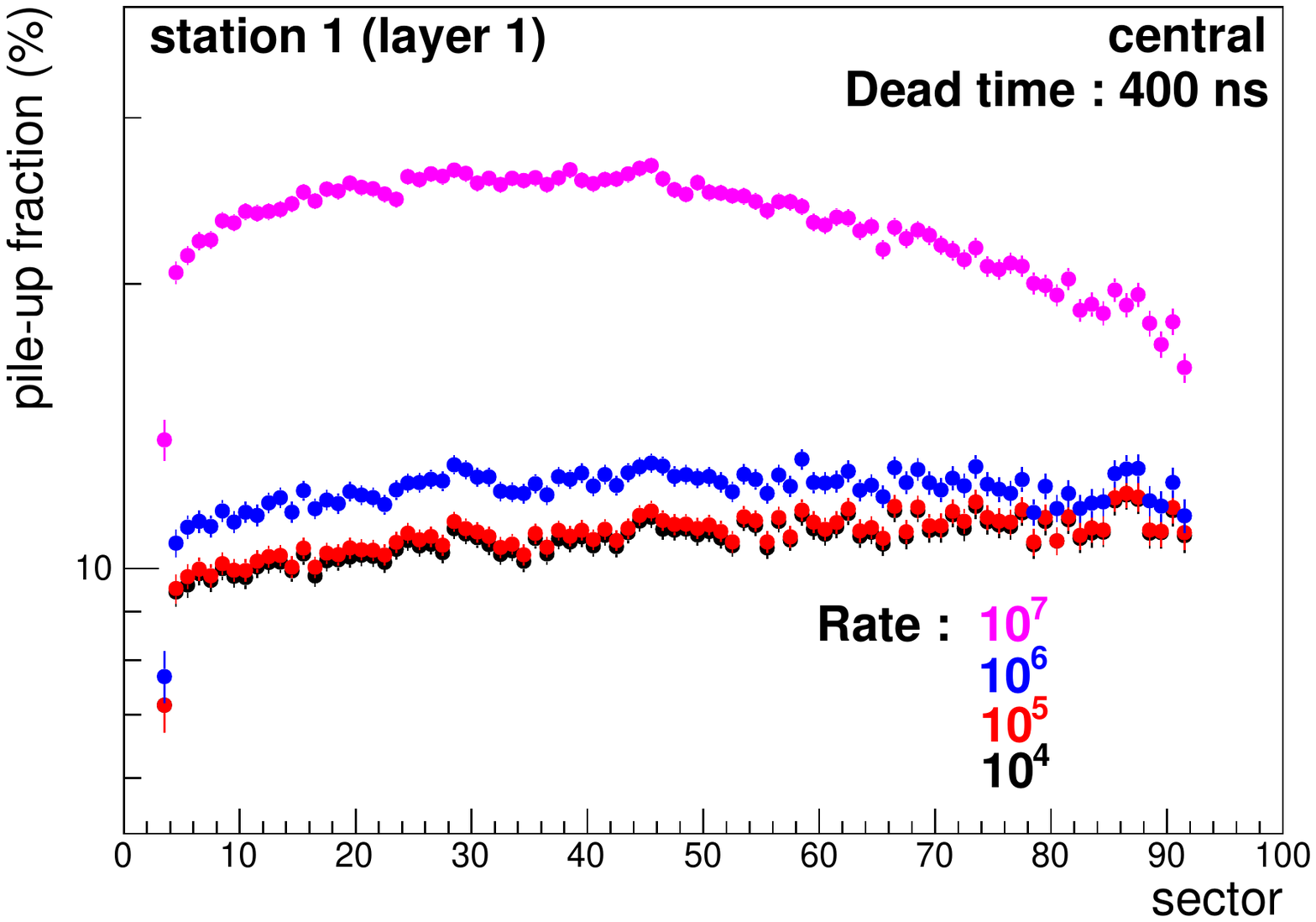}
\caption{Pile-up fraction in each sector of the first layer of the first MuCh station for a dead time of 400 ns and for different interaction rates. The left panel shows the (realistic) case of minimum-bias events, the right panel the (artificial) case of central events only.}
\label{fig:pileup_sector}
\end{center}
\end{figure}

The pile-up fraction is shown in Fig.~\ref{fig:pileup_sector} differentially for each sector in the first layer of the first station for different interaction rates, varying from $10^4$/s to $10^7$/s. Here one sector denotes one ring containing 360 pads which is according to progressive geometry (see right fig.~\ref{fig:much_geometry}). Again, the rate effect is better visible for central events (right-hand panel) than for minimum-bias events (left-hand panel). Pile-up at low rates happens between tracks within the same event and is thus irreducible by varying the dead time; it corresponds to conventional ''double hits‘‘ and is fixed by the granularity (pad layout) of the detector. From an event rate of about $10^6$/s on, the additional pile-up between tracks from different events becomes visible and increases with increasing rate.

\begin{figure}[htb]
\begin{center}
    \includegraphics[width=0.6\linewidth]{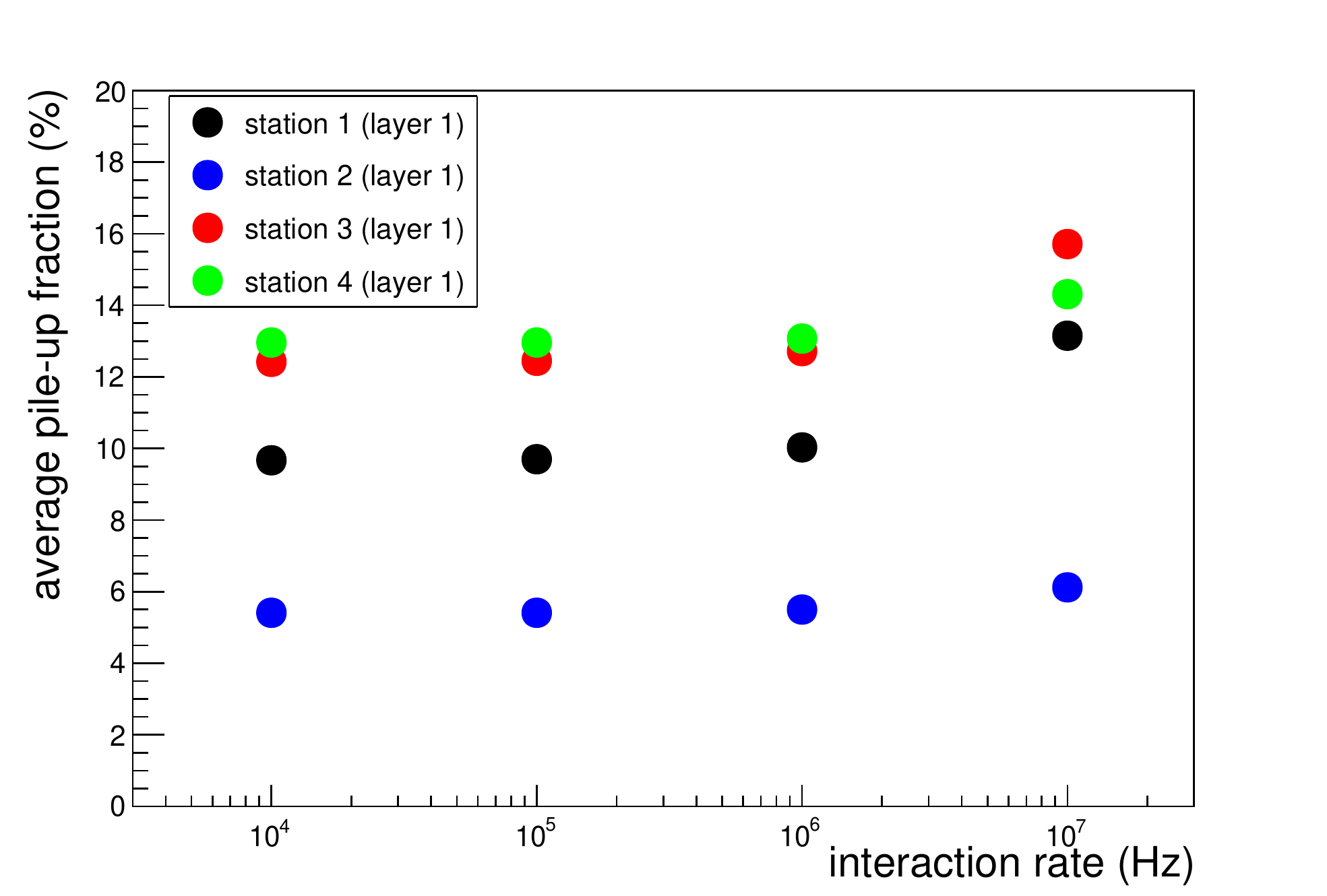}
\caption{Average pile-up fraction as a function of interaction rate for minimum-bias Au+Au collisions at 10$A$GeV/c. The single-channel dead time was taken to be 400 ns.}
\label{fig:pileup_station}
\end{center}
\end{figure}

\begin{figure}[htb]
\begin{center}
\includegraphics[width=0.49\linewidth]{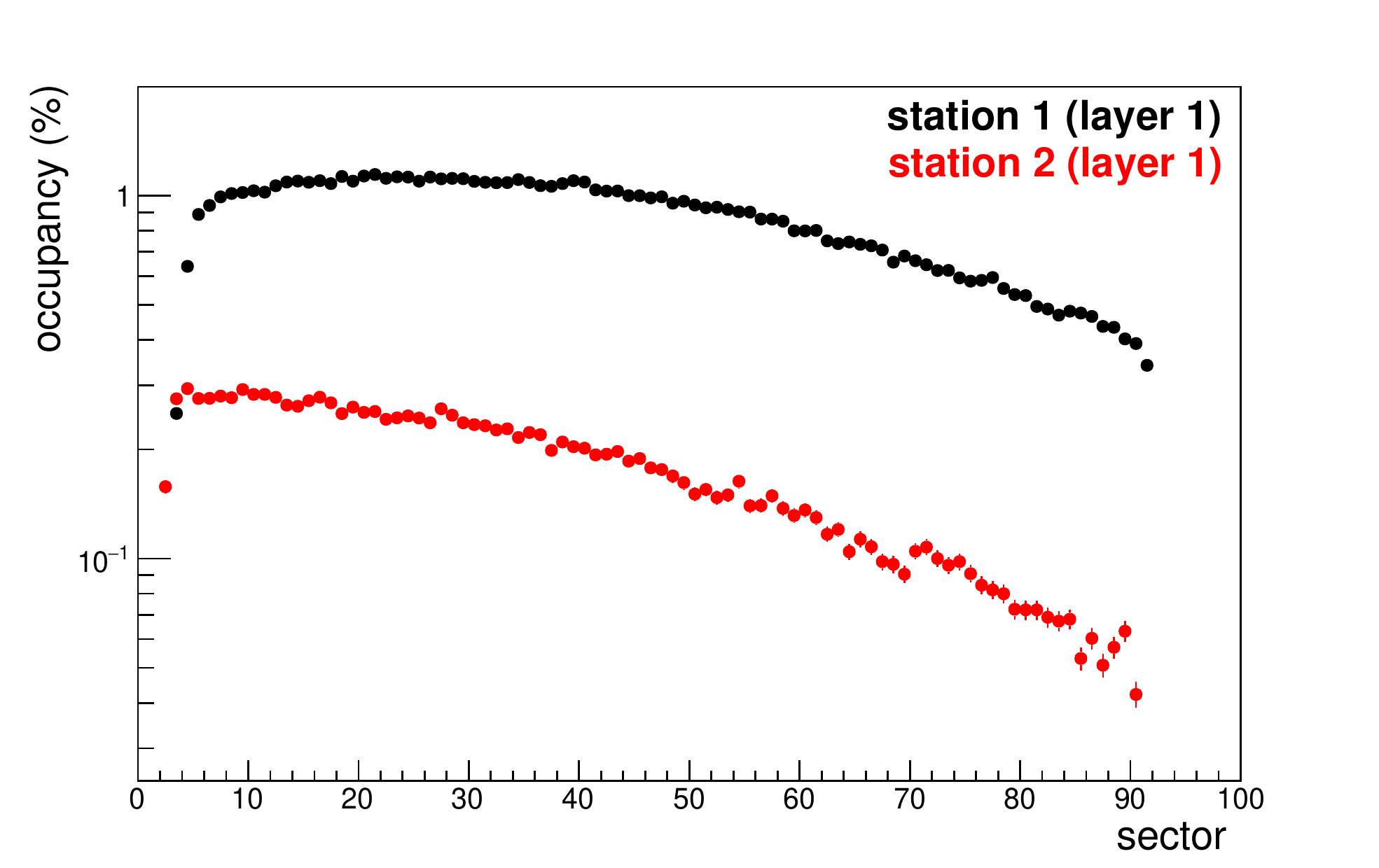}
\hfill
\includegraphics[width=0.47\linewidth]{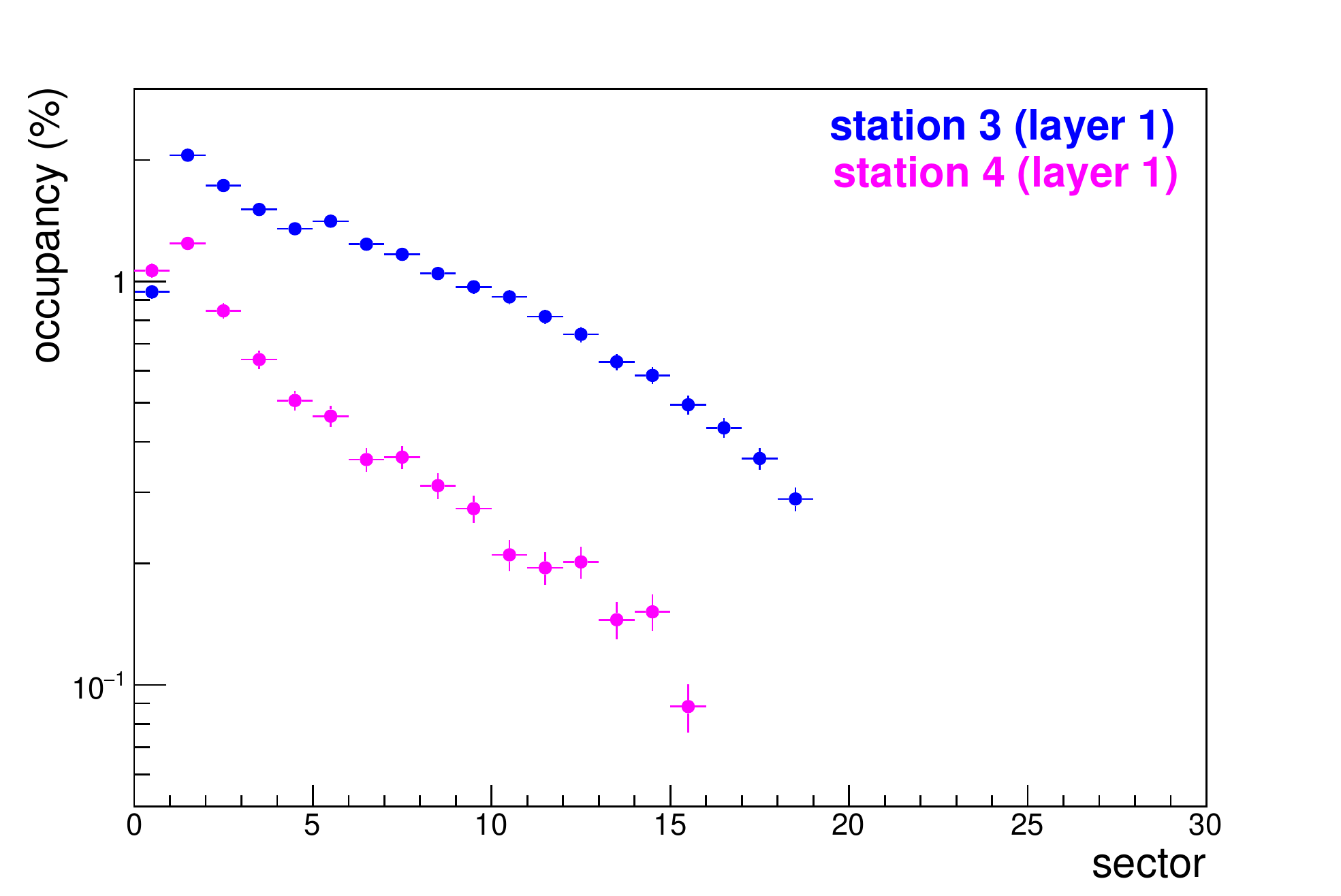}
\caption{Digi occupancy as function of sector of the first layer of $1^{st}$ and $2^{nd}$ MuCh station (left panel) and $3^{rd}$ and $4^{th}$ MuCh station (right panel).}
\label{fig:occupancy_sector}
\end{center}
\end{figure}

The situation in the other three MuCh stations is comparable to the one in the first station, as shown in Fig.~\ref{fig:pileup_station}. In all stations, an increase of inter-event pile-up with rate is observed on top of the irreducible in-event pile-up. Note that for constructional reasons, station 2 has the same angular pad segmentation as station 1, although the hit density is significantly reduced in the absorber segment in between which can be seen in left panel of fig.~\ref{fig:occupancy_sector}. This results in a lower occupancy and a lower amount of pile-up losses compared to station 1. The pad layout of stations 3 and 4 was chosen such as to result in an occupancy similar to station 1. The right panel of fig.~\ref{fig:occupancy_sector} shows the digi occupancy for first layer of $3^{rd}$ and $4^{th}$ MuCh station as function of sector. 

The consequence of the pile-up for the sensitivity of the detector to physics observables must be assessed by proper simulations including the full reconstruction of the simulated data stream in terms of hits and tracks. Such studies are yet to come. However, identifying the Monte-Carlo origin of the pile-up cases, we find that 92.47\% of all pile-ups happen between two secondary tracks in a shower created close to the downstream edge of an absorber layer. Such shower tracks are concentrated in a small emission cone are thus more likely to deliver charge into the same read-out pad. About 7.4\% of pile-ups occur between a primary and a secondary track; the fraction of pile-ups between two primary tracks (from the main event vertex) is negligible. This suggests the impact of pile-up on the physics performance of the detector to be minute.

\section{Summary}
\label{sec:summary}

We have presented a time-based simulation scheme for the MuCh detector of the CBM experiment which provides a realistic detector response of detector and read-out electronics in the environment of a trigger-less, free running data acquisition system. The key to this scheme is to enable interference of tracks from different events with small time separation, using an intelligent buffering procedure and a proper prescription for the treatment of signals arriving close in time in the same read-out channel (pad). The described MuCh simulation software is integrated into the common cbmroot simulation framework, which produces a data stream similar to that expected from the real experiment. The treatment of thermal, uncorrelated noise is included.

The simulation software allows to study the performance of the CBM-MuCh detector in a realistic scenario of a time sequence of minimum-bias nuclear collisions. In particular, rate-dependent effects like data losses arising from pile-up between subsequent events can be addressed, which is essential to assess the reconstruction efficiency of the detector under various operating conditions. Our findings of pile-up as functions of dead time and of the interaction rate are in qualitative agreement with our expectations, which to a certain extent verifies the software implementation. Quantitatively, we see significant losses from inter-event pile-up, above the irreducible in-event double hits, for interaction rates from $10^6$/s on. 

The developed software also allows to optimize the read-out of the detector in terms of parameters like single-channel dead time, time resolution or the applied threshold, thus providing feedback to the designers of the read-out ASIC SMX. The software will be validated against the real detector behaviour using data from laboratory and in-beam tests of detector prototypes. In particular, we plan to improve the currently simplified treatment of double hits in an ASIC channel, replacing the now used Theta function with the real signal shape in the slow channel of the SMX.

The quantitative effects of the rate-dependent data losses on the sensitivity of the detector to physics observables will be the subject of further studies. This requires the application of the full, time-based reconstruction (hit and track finding) on the generated data stream. The simulation software described in the article provides the required tools for such studies. 
\flushbottom

%\appendixpage
%\input{ClassDetails.tex}

\acknowledgments
We would like to express our sincere gratitude to the CBM Collaboration for all help. We are also very grateful to Dr. Anand Kumar Dubey, Mr. Jogender Saini, Ms Ekata Nandy and Mr. Ajit Kumar of VECC Kolkata for their valuable suggestions and various theoretical concepts during our work. We acknowledge the services and computing facility provided by the grid computing facility at VECC-Kolkata, India.

\end{document}